\documentstyle[amssymb,preprint,aps]{revtex}

\begin{document}
\title{Large Field Inflation in Supergravity}
\author{C. Panagiotakopoulos}
\address{Physics Division, School of Technology\\
Aristotle University of Thessaloniki\\
54006 Thessaloniki, Greece}
\maketitle

\begin{abstract}
We present a supergravity inflationary scenario in which the inflaton field
takes values considerably larger than the Planck scale. It is based on a
class of inflationary potentials which can be derived from ``singular'' K$%
\ddot{a}$hler potentials assuming simple superpotentials of the type $W\sim
S^n$. To this class belong, among many others, all potentials which are even
infinitesimally smaller than the one derived from the minimal K$\ddot{a} $%
hler potential. Our scenario allows for a detectable gravitational wave
contribution to the microwave background anisotropy.
\end{abstract}

\newpage

Most successful inflationary scenarios \cite{linde90} invoke a very weakly
coupled gauge singlet scalar field, the inflaton, in order to account for
the tiny temperature fluctuations $\frac{\Delta T}{T}$ in the cosmic
microwave background radiation. Nevertheless, the fine tuning that such a
weak coupling entails can be avoided in an ingenious model constructed by
Linde \cite{linde94} and studied in detail soon afterwards \cite{cope}.
Linde's model is a hybrid of chaotic inflation \cite{linde90} and the usual
theory of spontaneous symmetry breaking involving a possibly gauge
nonsinglet field. During inflation the non-inflaton field is trapped in a
false vacuum state and the universe is dominated by the false vacuum energy
density. Inflation ends with (or just before) a phase transition taking
place when the non-inflaton field rolls very rapidly to its true vacuum
state (``waterfall''). In the hybrid model the smallness of $\frac{\Delta T}{%
T}$ is not directly related to the smallness of the self-couplings of the
inflaton but can be obtained by exploiting the smallness of the false vacuum
energy density in Planck scale units together with an appropriate slope
along the inflationary trajectory. Thus, one has the option of forbidding
the inflaton self-couplings through appropriate symmetries. This is most
naturally implemented in the context of global supersymmetry by imposing
R-symmetries \cite{dvali}. Of course, one is still left with the problem of
generating the necessary slope along the inflationary trajectory. One
possibility is that this slope, or at least a significant part of it, is
generated when global supersymmetry is promoted to local \cite{pan1},\cite
{linde97},\cite{pan2}.

To investigate the consequences that supergravity has on hybrid inflationary
models we confine ourselves to the inflationary trajectory and use the
simple superpotential 
\begin{equation}
W=-\mu ^{2}S
\end{equation}
involving just the gauge singlet superfield $S$. $W$ is the most general
superpotential respecting the continuous R-symmetry $S\to e^{i\theta }S$, $%
W\to e^{i\theta }W$. In the context of global supersymmetry it gives rise to
a slopeless potential $V_{gl}=\mu ^{4}$ consisting entirely of the false
vacuum energy density $\mu ^{4}$ which plays the role of a ``temporary
cosmological constant''.

Let us now replace global supersymmetry by $N=1$ supergravity with a choice
of a minimal K$\ddot{a}$hler potential $K=\mid S\mid ^{2}$ leading to
canonical kinetic terms for the inflaton $\sigma $. [Through R-symmetry
transformations we bring the scalar component of the superfield $S$, for
which the same symbol $S$ is employed, to the form $S\equiv \frac{1}{\sqrt{2}%
}\sigma $, where $\sigma $ is a real scalar field. Throughout our discussion
we restrict ourselves to $\sigma \geqslant 0$ and we make use of units in
which the reduced Planck scale $m_{Pl}$ $\equiv \frac{M_{Pl}}{\sqrt{8\pi }}%
\simeq 2.4355\times 10^{18}\>GeV$ is equal to $1$ ($M_{Pl}\simeq 1.221\times
10^{19}\>GeV$ is the Planck mass).] Then, the ``canonical'' potential $%
V_{can}$ acquires a slope and becomes \cite{cope},\cite{pan1},\cite{linde97} 
\begin{equation}
V_{can}=V_{gl}(1-x+x^{2})e^{x}=\mu ^{4}(1-x+x^{2})e^{x}=\mu
^{4}\sum_{m=0}^{\infty }\frac{(m-1)^{2}}{m!}x^{m}\text{,}
\end{equation}
where $x\equiv \mid S\mid ^{2}=\frac{1}{2}\sigma ^{2}.$ $V_{can}$ of eq. (2)
does not allow inflation unless $x\ll 1$. From the expansion of $V_{can}$ as
a power series in $x$ we see that, due to an ``accidental'' cancellation,
the linear term in $x$ is missing and therefore no mass-squared term is
generated for $\sigma $.

Small deviations from the minimal form of the K$\ddot{a}$hler potential
respecting the R-symmetry lead to a K$\ddot{a}$hler potential \cite{pan2} 
\begin{equation}
K=x-\frac{\beta }{4}x^{2}+\cdots \text{.}
\end{equation}
The potential $V$ generated from such an almost-minimal K$\ddot{a}$hler
potential has an expansion in powers of $x$ of the form 
\begin{equation}
V=\mu ^{4}(1+\beta x+\cdots )
\end{equation}
in which a linear term proportional to the small parameter $\beta >0$ is now
generated. All higher powers of $x$ are still present in the series with
coefficients which are only slightly different from the corresponding ones
of eq. (2). In particular, there is a choice of the coefficients in eq. (3)
for which the resulting potential corresponds to the one of eq. (2) with
just the addition of the term $\beta x$ \cite{pan2}. Again we naively expect
inflation to be allowed only for $x\ll 1$.

Thus, one is tempted to conclude that the inflaton field variation in hybrid
inflation with canonical \cite{pan1},\cite{linde97} or quasi-canonical \cite
{pan2} supergravity is forced to be small in $m_{Pl}$ units and that such a
scenario necessarily allows only a limited number of e-foldings. Moreover,
as a consequence of the small inflaton field variation the gravitational
wave contribution to the cosmic microwave background anisotropy is expected
to be undetectable \cite{ly}. This conclusion is correct for canonical
supergravity because the K$\ddot{a}$hler potential is known exactly. For the
quasi-canonical case, however, one cannot safely decide by simply knowing a
few terms in the expansion of eq. (3).

It would certainly be very interesting if we could arrange for a scenario in
which the inflaton field takes values considerably larger than $m_{Pl}$
during the period of inflation relevant to the presently observable
universe. Naively, this has a chance to be achieved if the K$\ddot{a}$hler
potential differs substantially from the minimal one, such that the
resulting potential is a much more slowly increasing function than the
potential $V_{can}$ of eq. (2). It is then natural to expect the
coefficients in the expansion of eq. (3) to be all large and in particular
the coefficient $\beta $ of the first correction term to be of order unity
thereby forbidding inflation at small $\sigma $ values. Surprisingly enough
our naive expectations will prove wrong in the sense that, as we will
shortly see, even infinitesimally small deviations from the canonical
potential of eq. (2) could be sufficient to allow for an inflationary phase
above the Planck scale.

Before going into a more technical discussion let us briefly explain the
reason for this unexpected result. The important point is that when the
potential differs from the canonical one and consequently the K$\ddot{a}$%
hler potential is non-minimal the kinetic terms of the field $\sigma $
acquire ``corrections'' which affect its equation of motion. Because of
these complications the flatness of $V(\sigma )$ is not directly related to
inflation. To be able to easily decide whether inflation is allowed we
should find the canonically normalized field $\sigma _{infl}$ which obeys a
``conventional'' equation of motion and express the potential $V$ as a
function of it. It is the flatness of $V(\sigma _{infl})$ that is directly
related to inflation since the usual ``slow-roll'' parameters involve
derivatives of $V$ with respect to the canonically normalized field $\sigma
_{infl}$. It turns out that, even for small deviations from the canonical
potential, $\sigma _{infl}$ differs significantly from $\sigma $ at large
field values although it almost coincides with it at small field values.
Actually in the cases considered below $\sigma _{infl}$ diverges as $\sigma $
tends to a finite value $\sigma _{0}$ with $V(\sigma _{0})$ remaining
finite. Then, a tiny variation of $\sigma $ in the vicinity of $\sigma _{0}$
results to an infinite variation of $\sigma _{infl}$. Consequently, $%
V(\sigma _{infl})$ at large $\sigma _{infl}$ values is much flatter than $%
V(\sigma )$ in the vicinity of $\sigma _{0}.$ Thus, large field inflation in
our scheme is due to the fact that even infinitesimally small deviations
from the canonical potential, which are obviously unable to significantly
alter its value, are able to dramatically change the behavior of $\sigma
_{infl}(\sigma )$ leading to a function which blows up in the vicinity of a
finite point $\sigma _{0}$.

In order to achieve our goal we shall adopt the following procedure. We
start by choosing the potential $V(x)$ $>0$ (actually the supergravity
corrections $\frac{V}{V_{gl}}$ to the potential $V_{gl}$ of the globally
supersymmetric model) instead of the K$\ddot{a}$hler potential $K(x)$ and we
attempt to subsequently determine $K.$ Since the derivation of the potential 
$V$ from the K$\ddot{a}$hler potential $K$ involves partial differentiations
of $K\ $this inverse procedure is, in general, very difficult. In our case,
however, the determination of the K$\ddot{a}$hler potential is greatly
simplified by the fact that only the superfield $S$ is important during
inflation and by the existence of the R-symmetry which forces the K$\ddot{a}$%
hler potential to be a function of only one variable $x\equiv \mid S\mid
^{2}\ $instead of being a function of the two variables $S$ and $S^{*}$.
Then, assuming the linear superpotential $W=-\mu ^{2}S$ (and using the
relations\ $S\frac{\partial K}{\partial S}=S^{*}\frac{\partial K}{\partial
S^{*}}=x\frac{dK}{dx}$ and $\frac{\partial ^{2}K}{\partial S^{*}\partial S}=%
\frac{d}{dx}\left( x\frac{dK}{dx}\right) $), $V$ can be written in terms of $%
K$ in the form 
\begin{equation}
V=V_{gl}\left[ \left( 1+x\frac{dK}{dx}\right) ^{2}\left( \frac{d}{dx}\left( x%
\frac{dK}{dx}\right) \right) ^{-1}-3x\right] e^{K}
\end{equation}
which does not involve partial derivatives. This relation can be regarded as
an ordinary differential equation for $K(x)$%
\begin{equation}
\frac{d}{dx}\left( x\frac{dK}{dx}\right) =\left( 1+x\frac{dK}{dx}\right)
^{2}\left( 3x+\frac{V}{V_{gl}}e^{-K}\right) ^{-1}
\end{equation}
whose solution $K(x)$ satisfies the boundary conditions 
\begin{equation}
K\simeq x,\text{ }\frac{dK}{dx}\simeq 1,\text{ for }x\ll 1\text{.}
\end{equation}
The boundary conditions on $K$ do not really constrain $V$ since they simply
require that $V$ tends to $V_{gl}$ as $x$ tends to $0$. Thus, it seems that
in our case choosing the potential is equivalent to choosing the K$\ddot{a}$%
hler potential.

However, one does not have the freedom to further assume that the field $%
\sigma $, as a function of which $V$ is chosen, has canonical kinetic terms
and consequently obeys a ``conventional'' equation of motion because, as we
mentioned earlier, the kinetic terms in supergravity obtain ``corrections''
involving derivatives of the K$\ddot{a}$hler potential. For this reason it
is not easy to decide whether an input potential $V(x)$ allows inflation
without determining the K$\ddot{a}$hler potential. An equivalent description
is the one already adopted of determining the canonically normalized
candidate inflaton field $\sigma _{infl}$ which satisfies the differential
equation 
\begin{equation}
\frac{d}{d\sigma }\sigma _{infl}=\left[ \frac{d}{dx}\left( x\frac{dK}{dx}%
\right) \right] ^{\frac{1}{2}}=\left( 1+x\frac{dK}{dx}\right) \left( 3x+%
\frac{V}{V_{gl}}e^{-K}\right) ^{-\frac{1}{2}}
\end{equation}
with the boundary condition 
\begin{equation}
\sigma _{infl}\simeq \sigma ,\text{ for }x\ll 1\text{.}
\end{equation}
We see that determination of $\sigma _{infl}$ again necessitates
determination of $K$. The advantage of this description, however, is that $%
\sigma _{infl}$ obeys a ``conventional'' equation of motion and therefore
flatness of $V(\sigma _{infl})$ is sufficient for inflation to be allowed.

It should now be obvious that if the input function $V(x)$ is not singular
anywhere but the K$\ddot{a}$hler potential $K(x)$ and the canonically
normalized field $\sigma _{infl}(x)$ are singular at a finite point $x_{0}$,
then $V(\sigma _{infl})$ becomes flat as $\sigma _{infl}\rightarrow \infty $
(i.e. as $x\rightarrow x_{0}$). Indeed, the variation of $\sigma _{infl}(x)$
in a small interval $(x_{0}-\epsilon ,$ $x_{0}),$ with $0<\epsilon \ll 1,$
is infinite whereas, according to our assumption, the variation of $V(x)$ in
the same interval remains finite. This is the essence of our inflationary
scenario.

Such a scenario motivates us to assume the existence of potentials and K$%
\ddot{a}$hler potentials for which $\frac{V}{V_{gl}}e^{-K}$ sooner or later
tends to zero. Then, once $\frac{V}{V_{gl}}e^{-K}$ becomes negligible, the
above equations simplify considerably and the solutions follow the
asymptotic forms 
\begin{equation}
K=-3\ln \mid A-\ln x\mid -\ln x+B,
\end{equation}
\begin{equation}
\sigma _{infl}=-\sqrt{\frac{3}{2}}\ln \mid A-\ln x\mid +C,
\end{equation}
which obviously depend on $V$ only through the integration constants $A$, $B$
and $C$. Eq. (6) with $V(x)>0$ easily leads $($for $x>0)$ to $\frac{dK}{d\ln
x}=x\frac{dK}{dx}>0,$ from which it follows that $A\equiv \ln x+3\left( 1+x%
\frac{dK}{dx}\right) ^{-1}>\ln x$ when eqs. (10) and (11) start being
applicable. Both $K$ and $\sigma _{infl}$ tend to infinity when $x$ tends to
its largest allowed value $x_{0}=e^{A\text{ }}$with $V$ tending to the
finite, as we assume, value $V(x_{0})$. Then $V$, viewed as a function of $%
\sigma _{infl}$, soon becomes very flat thereby allowing for a very long
inflationary era. Indeed, for $x=x_{0}-\delta x$ (with $0<\delta x\ll 1$) $%
\frac{dV}{d\sigma _{infl}}=\frac{dV}{dx}\frac{dx}{d\sigma _{infl}}\simeq 
\sqrt{\frac{2}{3}}\frac{dV}{dx}\delta x$ which, for any finite value of $%
\frac{dV}{dx}\simeq \frac{dV}{dx}(x_{0})$, is as small as one wants provided 
$x$ is sufficiently close to $x_{0}$ or, equivalently, $\sigma _{infl}$
sufficiently large ($\sigma _{infl}\simeq \sqrt{\frac{3}{2}}\ln \frac{1}{%
\delta x}+\sqrt{\frac{3}{2}}A+C$ ). We see that our scenario is actually
consistent with eqs. (6) to (9) and consequently with supergravity. Whether
a given $V$ belongs to the above class of potentials which, as we will
shortly demonstrate, is non-empty can be easily tested numerically.

We first consider potentials $V$ which are only infinitesimally smaller than
the canonical potential $V_{can}$ 
\begin{equation}
V=V_{can}+\delta V,
\end{equation}
where $\delta V$ is negative and small. Equivalently, we consider potentials
whose series expansion in powers of $x$ has coefficients equal or slightly
smaller than the corresponding ones of eq. (2). Such potentials could arise
perturbatively from the minimal K$\ddot{a}$hler potential as a result of
small quantum corrections. It turns out that all potentials resulting from
such perturbations do belong to the desired class just described.

As an example, we consider the perturbation to $V_{can}$ given by $\delta
V/\mu ^{4}=-0.001x^{2}.$ In fig. 1 the numerical solution $K(\sigma )$ is
plotted. We see that $K(\sigma )$ practically coincides with the canonical
form $K(\sigma )=\frac{1}{2}\sigma ^{2}$ for $\sigma \lesssim 6,$ it remains
close to it until $\sigma $ approaches the largest allowed value $\sigma
_{0}\simeq 7.913$ and suddenly tends to infinity following eq. (10) as $%
\sigma $ tends to $\sigma _{0}$. The canonically normalized inflaton $\sigma
_{infl}(\sigma )$ is plotted as well. Again $\sigma _{infl}$ practically
coincides with $\sigma $ for $\sigma \lesssim 6,$ it remains close to it
until $\sigma $ approaches $\sigma _{0}$ and quickly follows the asymptotic
behavior of eq. (11) as $\sigma $ tends to $\sigma _{0}$. In fig. 2 the
potential $V(\sigma _{infl})$ is plotted. An enormous almost flat region for 
$\sigma _{infl}\gtrsim 12$ is apparent. Notice that this region corresponds
to a variation of $\sigma $ in the tiny interval $(\sigma _{0}-\varepsilon
,\sigma _{0})$ with $\varepsilon \simeq 0.017.$

The following arguments could help us ``understand'' this unexpected result.
Let us assume that at some point $x=x_{i}$ (with $x_{i}\ll 1$) $%
K(x_{i})=x_{i}$ and $\frac{dK}{dx}(x_{i})=1.$ Then, from $\frac{V}{V_{gl}}$ $%
<\frac{V_{can}}{V_{gl}}$ and eq. (5) one can easily show that $\frac{d^{2}K}{%
dx^{2}}(x_{i})>0.$ This means that there is an interval $(x_{i},\ x_{1})$
where $K(x)>x.$ In this interval $\frac{V}{V_{gl}}e^{-K}<\frac{V_{can}}{%
V_{gl}}e^{-x}=(1-x+x^{2})$ and consequently 
\begin{equation}
\left( 1+x\frac{dK}{dx}\right) ^{-2}\frac{d}{dx}\left( x\frac{dK}{dx}\right)
>\left( 1+x\right) ^{-2}.
\end{equation}
Integrating this relation from $x_{i}$ to $x$ (with $x_{i}<x<x_{1}$) we
obtain $\frac{dK}{dx}>1$ in the whole interval $(x_{i},\ x_{1})$ meaning
that the interval where $K(x)>x$ can be extended beyond $x_{1}.$ Repeating
this procedure we may conclude that $K(x)>x$ and $\frac{dK}{dx}>1$ hold for
all meaningful values of $x>x_{i}.$ Eq. (13), which now is assumed to hold
for all $x>x_{i}$, can be easily rewritten as 
\begin{equation}
\frac{d^{2}K}{dx^{2}}>x\left( \frac{dK}{dx}-1\right) \left( \frac{dK}{dx}-%
\frac{1}{x^{2}}\right) \left( 1+x\right) ^{-2},
\end{equation}
from which it follows that $\frac{dK}{dx}>1$ implies $\frac{d^{2}K}{dx^{2}}%
>0 $ for $x\geqslant 1.$ Thus, we expect that $\delta K\equiv K(x)-x,$ which
satisfies $\delta K>0,\ \frac{d\delta K}{dx}>0$ for $x>x_{i}$ and $\frac{%
d^{2}\delta K}{dx^{2}}>0$ for $x\geqslant 1$, will eventually grow
sufficiently fast forcing $\frac{V}{V_{gl}}e^{-K}<\frac{V_{can}}{V_{gl}}%
e^{-K}=(1-x+x^{2})e^{-\delta K}$ to tend to zero. Then, $K(x)$ and $\sigma
_{infl}(x)$ will be described by eqs. (10) and (11), respectively and $%
V(\sigma _{infl})$ will become asymptotically flat. Obviously, the
asymptotic flatness of $V(\sigma _{infl})$ could not be due directly to the
tiny perturbation $\delta V$ which is clearly unable to significantly alter
the value of $V_{can}$. The important effect of the small perturbation is
that it changes dramatically the behavior of $\sigma _{infl}(x)$ leading to
a function which blows up as $x$ approaches a finite value $x_{0}$ (with $%
V(x)$ remaining finite as $x\rightarrow x_{0}$). Notice that the whole
argument depends crucially on the boundary condition that the K$\ddot{a}$%
hler potential should tend to the minimal one at small field values.

It is worth emphasizing that the previous arguments do not involve the
magnitude of the (negative) perturbation which only determines the rate of
growth of $\delta K(x)$ with $x$ or, equivalently, the point $x_{0}$ in the
vicinity of which eqs. (10) and (11) start being applicable (the smaller $%
\left| \delta V\right| ,$ the larger $x_{0}$). Thus, we are led to the
important conclusion that all potentials $V$ with 
\begin{equation}
0<\frac{V}{V_{gl}}<\frac{V_{can}}{V_{gl}}
\end{equation}
allow inflation above the Planck scale.

The class of potentials allowing inflation above the Planck scale is
actually larger than the class of potentials resulting from the
perturbations just described. Thus, one could consider perturbations $\delta
V$ which are not necessarily negative for all values of $x$. For instance,
one could add to $V_{can}$ a small positive linear term in $x$ provided he
adds higher powers of $x$ with negative coefficients of appropriate
magnitude as well. Moreover, such perturbations do not have to be
necessarily small. Thus, there are potentials which could be regarded as
large deviations from $V_{can}$ which still belong to the desired class.

As an illustration we consider potentials of the type 
\begin{equation}
V=\mu ^{4}\left[ 1+(\beta -\gamma )x+\alpha x^{2}\right] e^{\gamma x}=\mu
^{4}(1+\beta x+\cdots )\text{ }
\end{equation}
involving the three (non-negative) real parameters $\alpha $, $\beta $ and $%
\gamma $ in addition to the false vacuum energy density $\mu ^{4}$. The
choice $\alpha =\gamma =1$, $\beta =0$ corresponds to $V_{can}$. For $\beta
>0$ all such potentials are larger than $V_{can}$ for sufficiently small
values of $x.$ Such potentials with $\alpha =1$, $\gamma \leqslant 1-\beta $
or $\alpha =\gamma \leqslant 1-\frac{\beta }{2}$ do belong to the desired
class not only for $\beta \ll 1$ but also for larger values of $\beta .$
Moreover, this class contains potentials with $\alpha =\beta =1,\ \gamma
\lesssim 0.565$ or $\alpha =\beta =\gamma \lesssim 0.738.$ Even more
surprisingly it contains potentials with $\beta =\gamma \sim 1$ provided $%
\alpha $ is sufficiently small$.$ More specifically, $\beta =\gamma =1$ is
allowed provided $\alpha \lesssim 0.187$ or $\beta =\gamma $ could be as
large as $1.253$ if $\alpha =0$. Clearly, the allowed values of $\alpha $
and $\gamma \ $are not determined sharply by the value of $\beta $ but
instead they belong to quite large regions in the $\alpha -\gamma $ plane.
Table 1 gives the values of the integration constants $A$, $B$ and $C$
appearing in the asymptotic solutions for $K$ and $\sigma _{infl}$ for
several values of the parameters $\alpha $, $\beta $ and $\gamma $ which
give rise to potentials belonging to the class we are interested in. In fig.
3 the numerical solutions for $K(\sigma )$ and $\sigma _{infl}(\sigma )$ are
plotted for the choice of parameters $\alpha =1$, $\beta =1$, $\gamma =0.5$.
Deviations from the corresponding canonical forms are now certainly larger
than the ones in fig. 1. Still, the potential $V(\sigma _{infl})$ is flat
for $\sigma _{infl}\gtrsim 8$ as seen from fig. 4.

A potential of the type $V=\mu ^{4}e^{\frac{\delta }{6}\sigma ^{2}}$ (with $%
0<\frac{\delta }{3}\ll 1$) along the inflationary trajectory in supergravity
has been suggested earlier in connection with ``tilted hybrid inflation'' 
\cite{bel}. This is our potential of eq. (16) with $\alpha =0$,$\ \beta
=\gamma =\frac{\delta }{3}\ll 1$. However, such a potential has not been
shown in \cite{bel} to be derivable from a K$\ddot{a}$hler potential and for
this reason the fact that the canonically normalized inflaton really differs
from $\sigma $ has been overlooked. Consequently, inflation seemed forbidden
for large $\sigma $ values which had as a result a limited total number of
e-foldings. Even if $\beta =\gamma =\frac{\delta }{3}$ is chosen
sufficiently small and inflation is naively expected to take place at
relatively small $\sigma $ values, the non-canonical kinetic terms could
play an important role if strong radiative corrections \cite{dvali} to the
slope of the inflationary trajectory are present. In our opinion, as we
already emphasized, small values simultaneously for $\alpha ,$ $\beta $ and $%
\gamma $ are rather unnatural because they suggest a small first correction
to the canonical K$\ddot{a}$hler potential with all the higher ones in the
expansion of eq. (3) being rather large. From our earlier discussion follows
that $\delta $ could be quite large  ($\frac{\delta }{3}$ $\lesssim 1.253$).
Then inflation, which is now forbidden at small $\sigma $, must necessarily
take place at $\sigma $ values close to the largest allowed value $\sigma
_{0}$.

Barring the unnatural situation of a very flat potential with$\ \beta \ll 1,$
inflation for $x\gtrsim 1$ is not allowed unless $\frac{d}{dx}\sigma _{infl}$
becomes large i.e. unless $\sigma _{infl}$ is described by eq. (11) with $x$
close to $x_{0}=e^{A}$. Then, the number of e-foldings $\Delta
N(x_{in},x_{f})$ for the time period that $x$ varies between the values $%
x_{in}$ and $x_{f}$ ($x_{in}\geqslant x_{f}$) is given, in the slow roll
approximation, by 
\begin{equation}
\Delta N(x_{in},x_{f})=-\int_{x_{in}}^{x_{f}}V\left( \frac{dV}{dx}\right)
^{-1}\left( \frac{d}{dx}\sigma _{infl}\right) ^{2}dx\simeq \frac{3}{2}\left( 
\frac{d\ln V}{d\ln x}\right) _{x_{0}}^{-1}\left( A-\ln x\right) ^{-1}\mid
_{x_{f}}^{x_{in}}.
\end{equation}
With $x_{H}$ being the value of $x$ when the scale $\ell _{H}$,
corresponding to the present horizon, crossed outside the inflationary
horizon and $x_{end}$ its value at the end of inflation, $N_{H}\equiv \Delta
N(x_{H},x_{end})$ is estimated to be 
\begin{equation}
N_{H}\simeq \frac{3}{2}\left( \frac{d\ln V}{d\ln x}\right)
_{x_{0}}^{-1}\left( A-\ln x_{H}\right) ^{-1}.
\end{equation}
Using this relation we obtain estimates for the slow-roll parameters $\frac{%
V^{\prime }}{V}$ and $\frac{V^{\prime \prime }}{V}$ (where the prime refers
to differentiation with respect to $\sigma _{infl}$) at the scale $\ell _{H}$
\begin{equation}
\left( \frac{V^{\prime }}{V}\right) _{x_{H}}\simeq \sqrt{\frac{3}{2}}%
N_{H}^{-1},\quad \left( \frac{V^{\prime \prime }}{V}\right) _{x_{H}}\simeq
-N_{H}^{-1}
\end{equation}
and for the differential spectral index $n_{H}$%
\begin{equation}
n_{H}\simeq 1+2\left( \frac{V^{\prime \prime }}{V}\right) _{x_{H}}-3\left( 
\frac{V^{\prime }}{V}\right) _{x_{H}}^{2}\simeq 1-2N_{H}^{-1}.
\end{equation}

For the quadrupole anisotropy $\frac{\Delta T}{T}$ we employ the standard
formula \cite{lid} 
\begin{equation}
\left( \frac{\Delta T}{T}\right) ^{2}\simeq \frac{1}{720\pi ^{2}}\left[ 
\frac{V^{3}}{V^{\prime 2}}+6.9V\right] _{x_{H}},
\end{equation}
from which the parameter $\mu $ is estimated 
\begin{equation}
\mu \simeq \left( 1080\pi ^{2}\right) ^{\frac{1}{4}}\left(
N_{H}^{2}+10.35\right) ^{-\frac{1}{4}}\left( \frac{V}{\mu ^{4}}\right)
_{x_{0}}^{-\frac{1}{4}}\left( \frac{\Delta T}{T}\right) ^{\frac{1}{2}}.
\end{equation}
The first term in eq. (21) is the scalar component $(\frac{\Delta T}{T}%
)_{S}^{2}$ of $(\frac{\Delta T}{T})^{2}$ whereas the second is the tensor
one $(\frac{\Delta T}{T})_{T}^{2}$ which represents the gravitational wave
contribution. Their ratio $r$ is 
\begin{equation}
r\equiv \left( \frac{\Delta T}{T}\right) _{T}^{2}/\left( \frac{\Delta T}{T}%
\right) _{S}^{2}\simeq 6.9\left( \frac{V^{\prime }}{V}\right)
_{x_{H}}^{2}\simeq 10.35N_{H}^{-2}\text{.}
\end{equation}

Taking $N_{H}\simeq 60$ in the above formulas we obtain $n_{H}\simeq 0.97$
and $r\simeq 3\times 10^{-3}$. Therefore, the gravitational wave signal is
undetectably small \cite{tur} in this scheme.

Our inflationary scenario, which is based on ``singular'' K$\ddot{a}$hler
potentials, resembles the scenario considered in \cite{st}. An important
difference between the two is that in our case the superpotential during
inflation is the typical one encountered in models of false vacuum inflation
whereas in \cite{st} conditions are imposed on the superpotential which are
not satisfied in the simplest models.

The above formulas suggest that if the value $N_{H}\simeq 12$ is employed
then $r\simeq 7\times 10^{-2}$ could be obtained leading possibly to a
detectable gravitational wave signal \cite{tur}. Of course, such a low value
of $N_{H}$ is allowed only if the inflationary stage just described is
followed by a second one at values of $x\ll 1$ producing the additional
number of e-foldings necessary for the solution of the cosmological
problems. This, in turn, becomes possible on the same inflationary
trajectory provided $\beta \ll 1$ and the potential is not too steep for $%
0.1\lesssim x\lesssim x_{0}$ such that a second stage of inflation
complementary to the first one does take place. An example of such a
potential is 
\begin{equation}
V=V_{can}+\delta V
\end{equation}
with 
\begin{equation}
\frac{\delta V}{\mu ^{4}}=f(x)=\beta
x-0.04x^{2}-0.2x^{3}-0.245x^{4}-0.13x^{5}-0.034x^{6}-0.007x^{7}-0.001x^{8}
\end{equation}
and $0.03\lesssim \beta \lesssim 0.035.$ It is derived from a K$\ddot{a}$%
hler potential $K$ which for $x\ll 1$ admits an expansion in powers of $x$ 
\begin{equation}
K=\sum_{n=1}^{\infty }a_{n}x^{n}
\end{equation}
with $\left| a_{n}\right| \lesssim 10^{-2}$ for $n\neq 1.$ The values of the
first few $a_{n}$ 's are given in table 2. Fig. 5 gives the plots of $%
K(\sigma )$ and $\sigma _{infl}(\sigma )$ whereas fig. 6 gives the plot of $%
V(\sigma _{infl}).$ Everywhere the choice $\beta =0.03$ was made.

Our discussion so far assumes a linear superpotential $W=-\mu ^{2}S$ leading
to a potential which tends to a constant $\mu ^{4}$ at small field values.
Consequently, during inflation the universe is trapped in a false vacuum
state and the inflationary stage has to be followed by a phase transition in
order for the false vacuum energy density $\mu ^{4}$ to be cancelled. Our
results can be easily extended to the case of inflation taking place with
the universe being in its true vacuum state with a superpotential 
\begin{equation}
W=\frac{\lambda }{n}S^{n},\ \ n>1
\end{equation}
respecting the continuous R-symmetry $S\rightarrow e^{i\theta }S,\
W\rightarrow e^{in\theta }W.$ In the context of global supersymmetry the
superpotential of eq. (27) gives rise to a potential 
\begin{equation}
V_{gl}=\left| \lambda \right| ^{2}\left| S\right| ^{2n-2}=\left| \lambda
\right| ^{2}x^{n-1}
\end{equation}
leading to the usual chaotic inflation \cite{linde90} for $x\gg 1$ . In N=1
supergravity the potential becomes 
\begin{equation}
V=V_{gl}\left[ \left( 1+\frac{x}{n}\frac{dK}{dx}\right) ^{2}\left( \frac{d}{%
dx}\left( x\frac{dK}{dx}\right) \right) ^{-1}-\frac{3}{n^{2}}x\right] e^{K},
\end{equation}
which for the minimal K$\ddot{a}$hler potential $K(x)=x$ gives 
\begin{equation}
V_{can}=V_{gl}(1+\frac{2n-3}{n^{2}}x+\frac{1}{n^{2}}x^{2})e^{x}=V_{gl}%
\sum_{m=0}^{\infty }\frac{(n+m)^{2}-4m}{n^{2}}\frac{1}{m!}x^{m}
\end{equation}
forbidding inflation for all values of $x.$ However, there are again
non-singular potentials, including all potentials satisfying $0<\frac{V}{%
V_{gl}}<\frac{V_{can}}{V_{gl}}$, for which $\frac{V}{V_{gl}}e^{-K}$ tends to
zero as $x$ tends to a finite point $x_{0}$ leading to diverging $K(x)$ and $%
\sigma _{infl}(x)$ and consequently to a large field inflationary scenario.
In particular eq. (11) remains unaltered as do eqs. (17) to (23) (with the
exception of eq. (22) in which $\mu $ should be replaced by $\sqrt{\left|
\lambda \right| }$). Of course, with $n>1$ inflation in two stages on the
same trajectory is not possible since, as is well known, $V_{gl}$ of eq.
(28) does not allow inflation at small field values.

To summarize, we presented an inflationary scenario in supergravity taking
place at inflaton field values considerably larger than $m_{Pl}$ and based
on potentials derived from ``singular'' K$\ddot{a}$hler potentials. Of
particular interest are cases in which the K$\ddot{a}$hler potential is very
close to the minimal one for small inflaton field values. In some cases our
scenario predicts a detectable gravitational wave signal in the cosmic
microwave background anisotropy. Whether K$\ddot{a}$hler potentials similar
to the ones required for the realization of our inflationary scenario are
obtainable in the context of more fundamental constructions giving
effectively rise to supergravity remains a challenging open issue.

\acknowledgments

This research was supported in part by EU under TMR contract
ERBFMRX-CT96-0090 and under HCM contract CHRX-CT94-0423. The author would
like to thank N. Ganoulis and G. Lazarides for useful discussions.

\bigskip $
\begin{array}{cccccc}
\alpha & \beta & \gamma & A & B & C \\ 
&  &  &  &  &  \\ 
0 & 1 & 0 & 1.70765 & 2.60533 & 1.97641 \\ 
1 & 1 & 0.1 & 1.95456 & 3.22173 & 2.08914 \\ 
1 & 1 & 0.3 & 2.11480 & 3.78278 & 2.19302 \\ 
1 & 1 & 0.5 & 2.52255 & 5.82376 & 2.55717 \\ 
0.738 & 0.738 & 0.738 & 3.48979 & 20.80807 & 4.64869 \\ 
0 & 0.5 & 0.5 & 1.61063 & 2.47728 & 1.95865 \\ 
0 & 0.75 & 0.75 & 1.72307 & 2.72645 & 2.00278 \\ 
0 & 1 & 1 & 1.90022 & 3.24982 & 2.09902 \\ 
1 & 0.03 & 0.97 & 2.87459 & 13.93121 & 3.77610 \\ 
1 & 0.05 & 0.95 & 2.74657 & 11.78929 & 3.48537 \\ 
1 & 0.07 & 0.93 & 2.65533 & 10.40770 & 3.28897 \\ 
0.985 & 0.03 & 0.985 & 3.04168 & 17.09767 & 4.18074 \\ 
0.975 & 0.05 & 0.975 & 2.92584 & 14.82190 & 3.89277 \\ 
0.965 & 0.07 & 0.965 & 2.84305 & 13.33466 & 3.69675
\end{array}
$

\qquad

\medskip

\medskip

Table 1. The values of the integration constants $A$, $B$ and $C$ for
various values of the parameters $\alpha $, $\beta $ and $\gamma $ in the
potential of eq. (16). \newpage

$
\begin{array}{cc}
\qquad n\qquad & \qquad a_{n}\qquad \\ 
&  \\ 
1 & +1. \\ 
2 & -0.007500 \\ 
3 & +0.010378 \\ 
4 & +0.002090 \\ 
5 & -0.001044 \\ 
6 & +0.000705 \\ 
7 & -0.000401 \\ 
8 & +0.000278 \\ 
9 & -0.000221 \\ 
10 & +0.000188
\end{array}
$

\bigskip

Table 2. The values of the first few coefficients appearing in the expansion
of the K$\ddot{a}$hler potential $K$ of eq. (26).

\bigskip \newpage

\smallskip Fig. 1. The K$\ddot{a}$hler potential $K$ and the canonically
normalized inflaton $\sigma _{infl}$ are plotted as functions of $\sigma $
for the choice $\delta V/\mu ^{4}=-0.001x^{2}$ of the perturbation to $%
V_{can}$. Their ``canonical'' forms are plotted as well.

Fig. 2. The potential $V$ is plotted as a function of the canonically
normalized inflaton $\sigma _{infl}$ for the choice $\delta V/\mu
^{4}=-0.001x^{2}$ of the perturbation to $V_{can}$.

Fig. 3. The K$\ddot{a}$hler potential $K$ and the canonically normalized
inflaton $\sigma _{infl}$ are plotted as functions of $\sigma $ for the
choice $\alpha =1,\beta =1$ and $\gamma =0.5$ of the parameters in the
potential of eq. (16). Their ``canonical'' forms are plotted as well.

Fig. 4. The potential $V$ is plotted for the choice $\alpha =1,\beta =1$ and 
$\gamma =0.5$ of the parameters in the potential of eq. (16) as a function
of the canonically normalized inflaton $\sigma _{infl}$.

Fig. 5. The K$\ddot{a}$hler potential $K$ and the canonically normalized
inflaton $\sigma _{infl}$ are plotted as functions of $\sigma $ for the
choice $\delta V/\mu ^{4}=f(x)$ of the perturbation to $V_{can}$. Their
``canonical'' forms are plotted as well.

Fig. 6. The potential $V$ is plotted as a function of the canonically
normalized inflaton $\sigma _{infl}$ for the choice $\delta V/\mu ^{4}=f(x)$
of the perturbation to $V_{can}$.

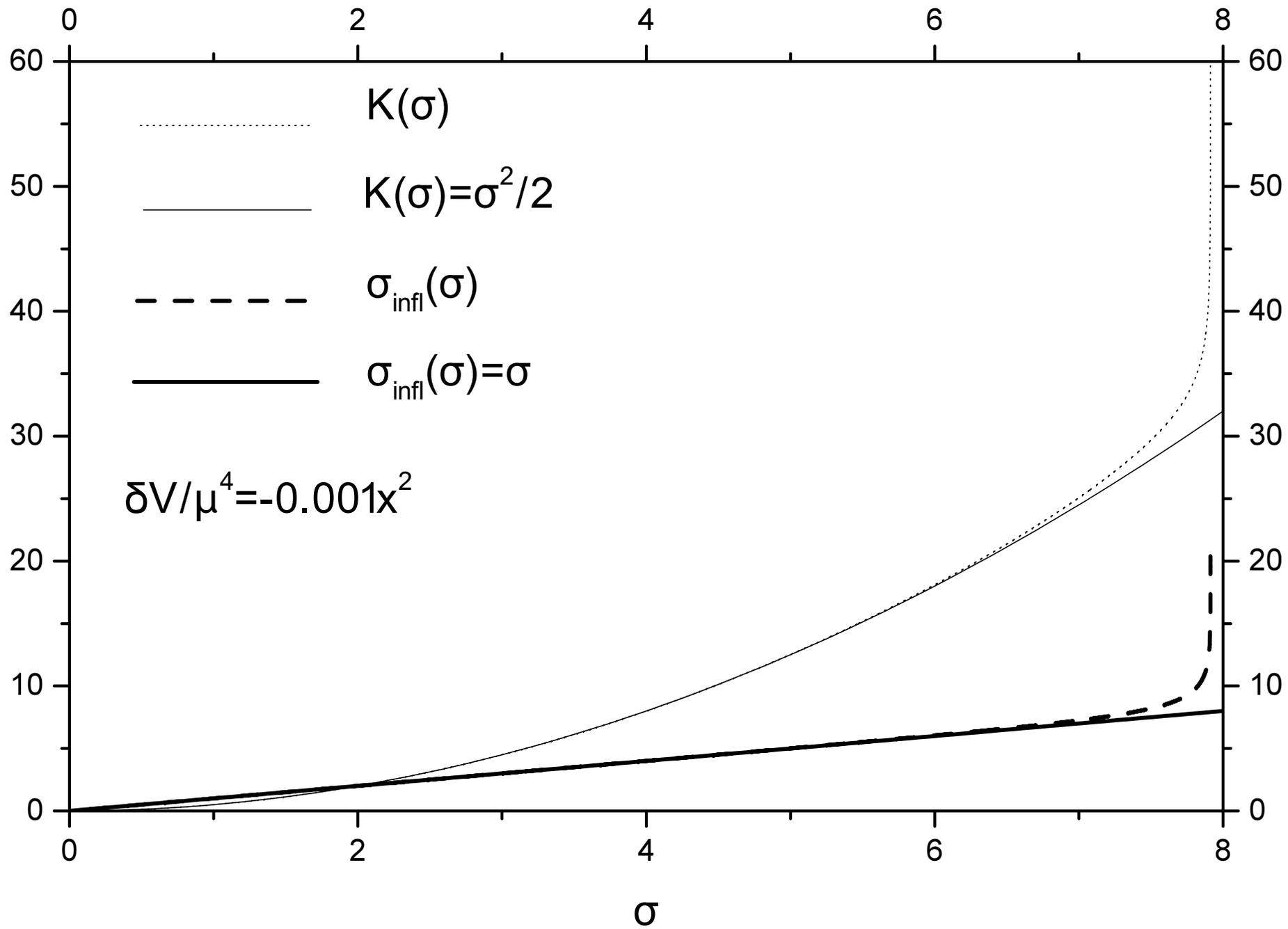
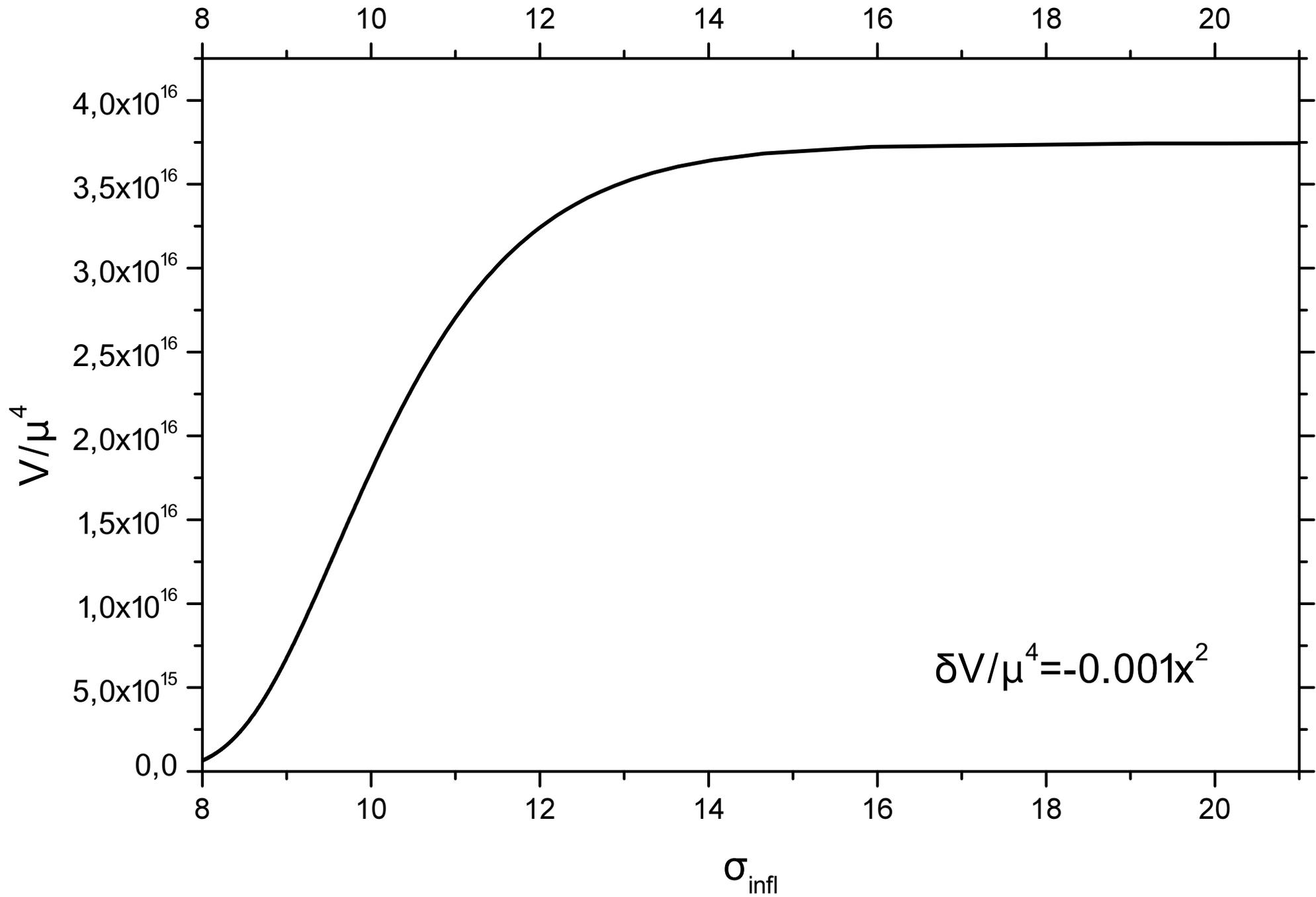
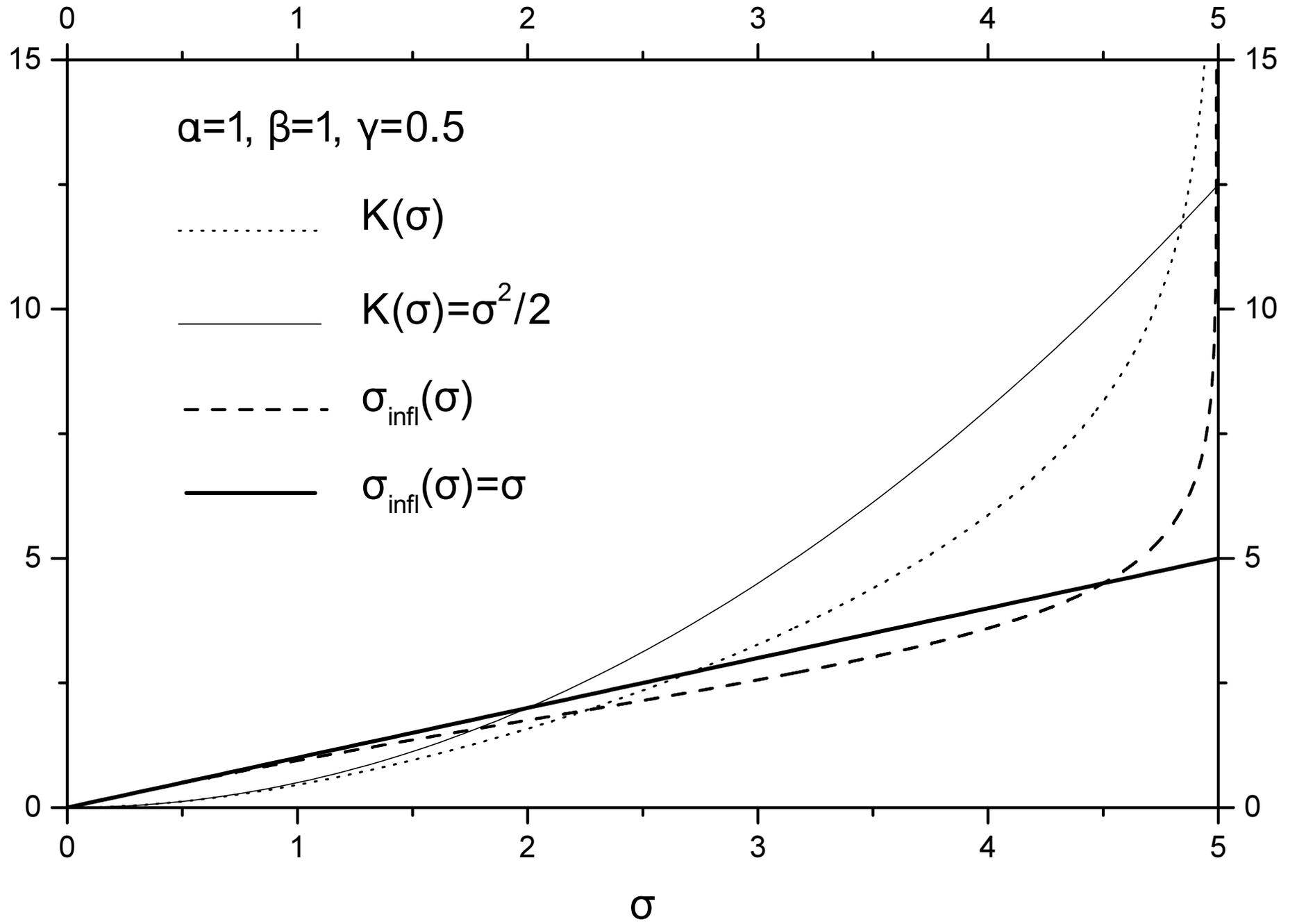
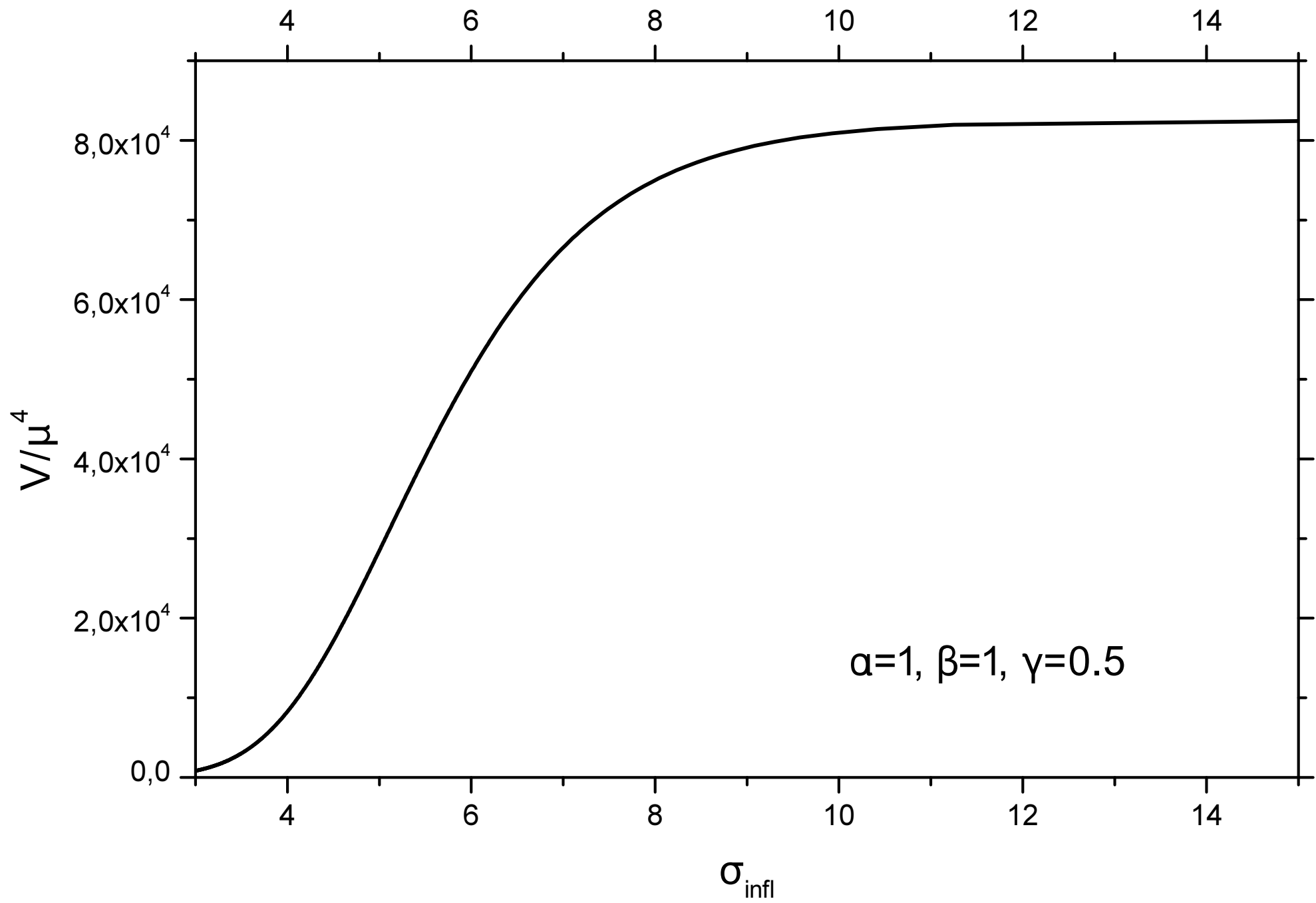
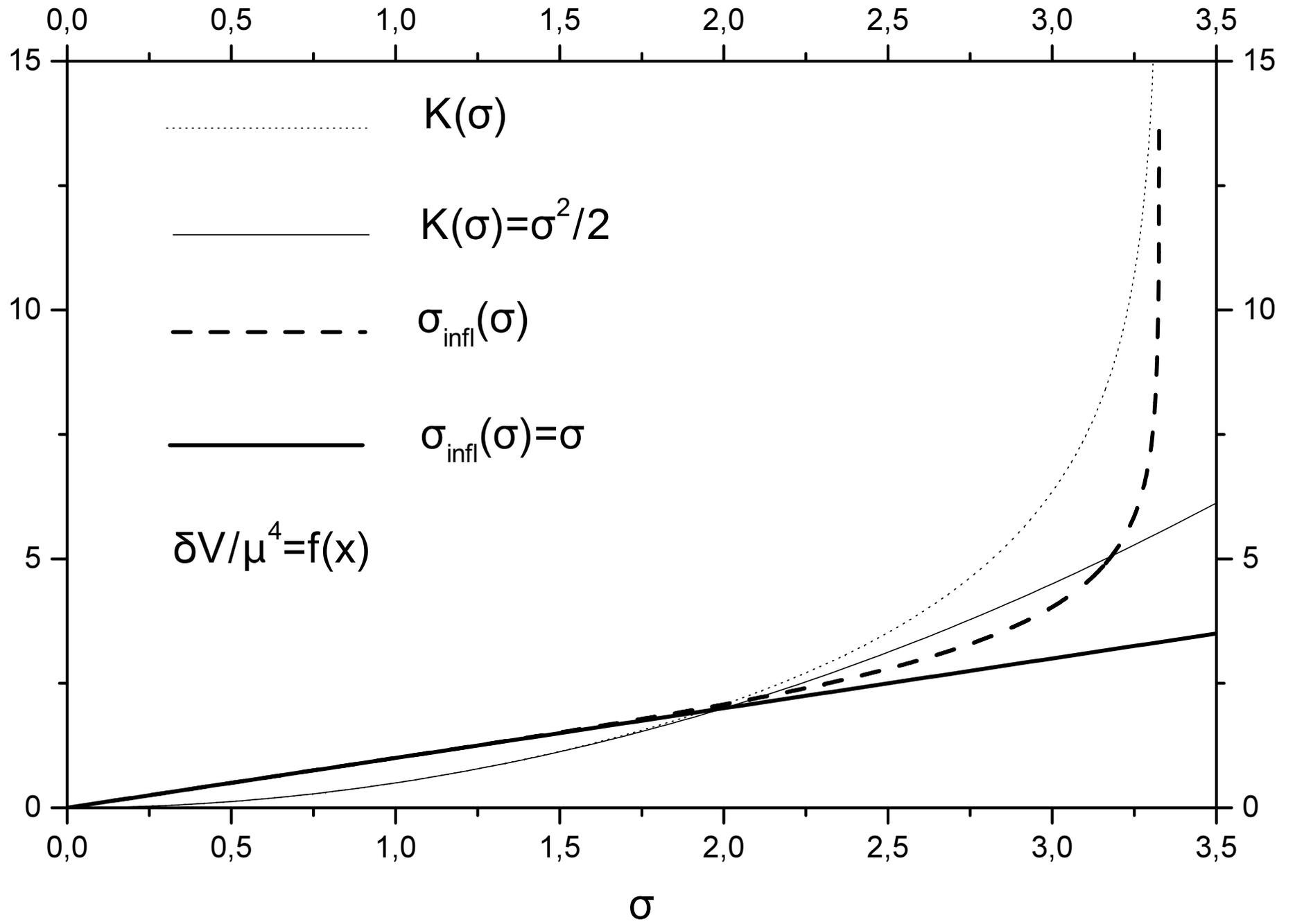
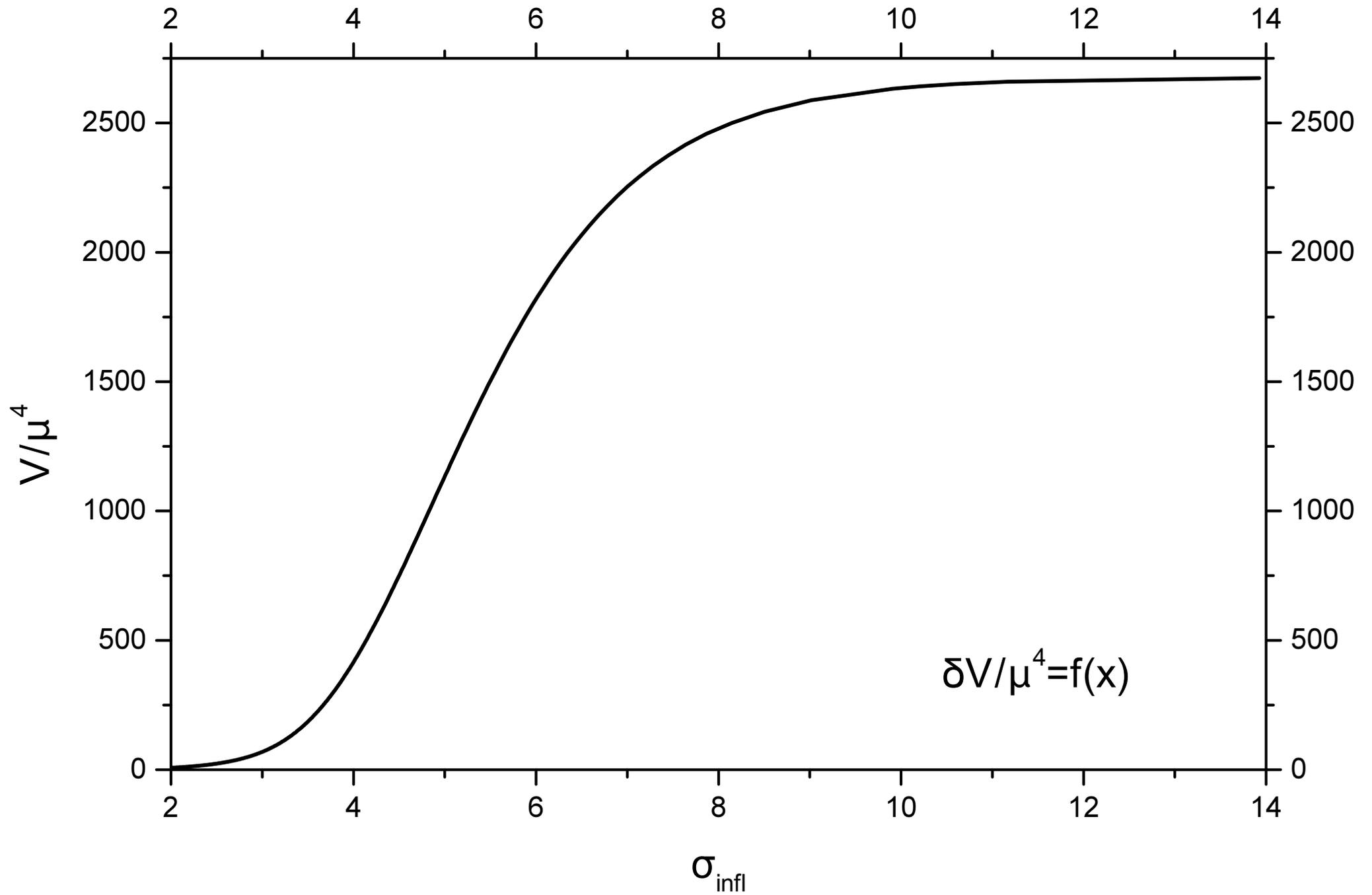
\end{document}